\documentclass{vgtc}                          % final (conference style)

\ifpdf%                                
  \pdfoutput=1\relax                   
  \usepackage{graphicx}                
  \DeclareGraphicsExtensions{.pdf,.png,.jpg,.jpeg} 
\else%                                 
  \usepackage{graphicx}                
  \DeclareGraphicsExtensions{.eps}     
\fi%

\graphicspath{{figures/}{pictures/}{images/}{./}} 

\usepackage{microtype}                 
\PassOptionsToPackage{warn}{textcomp}  
\usepackage{textcomp}                  
\usepackage{mathptmx}                  
\usepackage{times}                     
         
\usepackage{cite}                      
\usepackage{tabu}                      
\usepackage{booktabs}                  

\onlineid{1146}

\vgtccategory{Research}

\usepackage{xcolor}
\usepackage[numbers]{natbib}
\usepackage{booktabs}
\usepackage{multirow}
\usepackage{url} 

\newcommand{\toolName}{\textit{Intercept Graph}}

\usepackage[linewidth=0pt,linecolor=blue]{mdframed}
\newcommand{\modify}[1]{\textcolor{black}{#1}}
\newcommand{\modifY}[1]{\textcolor{black}{#1}}
\title{{\toolName}: An Interactive Radial Visualization for Comparison of State Changes}
\author{Shaolun Ruan\thanks{e-mail: haywardryan@foxmail.com}\\ %
        \scriptsize Kent State University %
\and Yong Wang\thanks{e-mail: yongwang@smu.edu.sg}\\ %
     \scriptsize Singapore Management University %
\and Qiang Guan\thanks{e-mail: qguan@kent.edu}\\ %
     \scriptsize Kent State University}

\teaser{
  \centering
  \includegraphics[width=\linewidth]{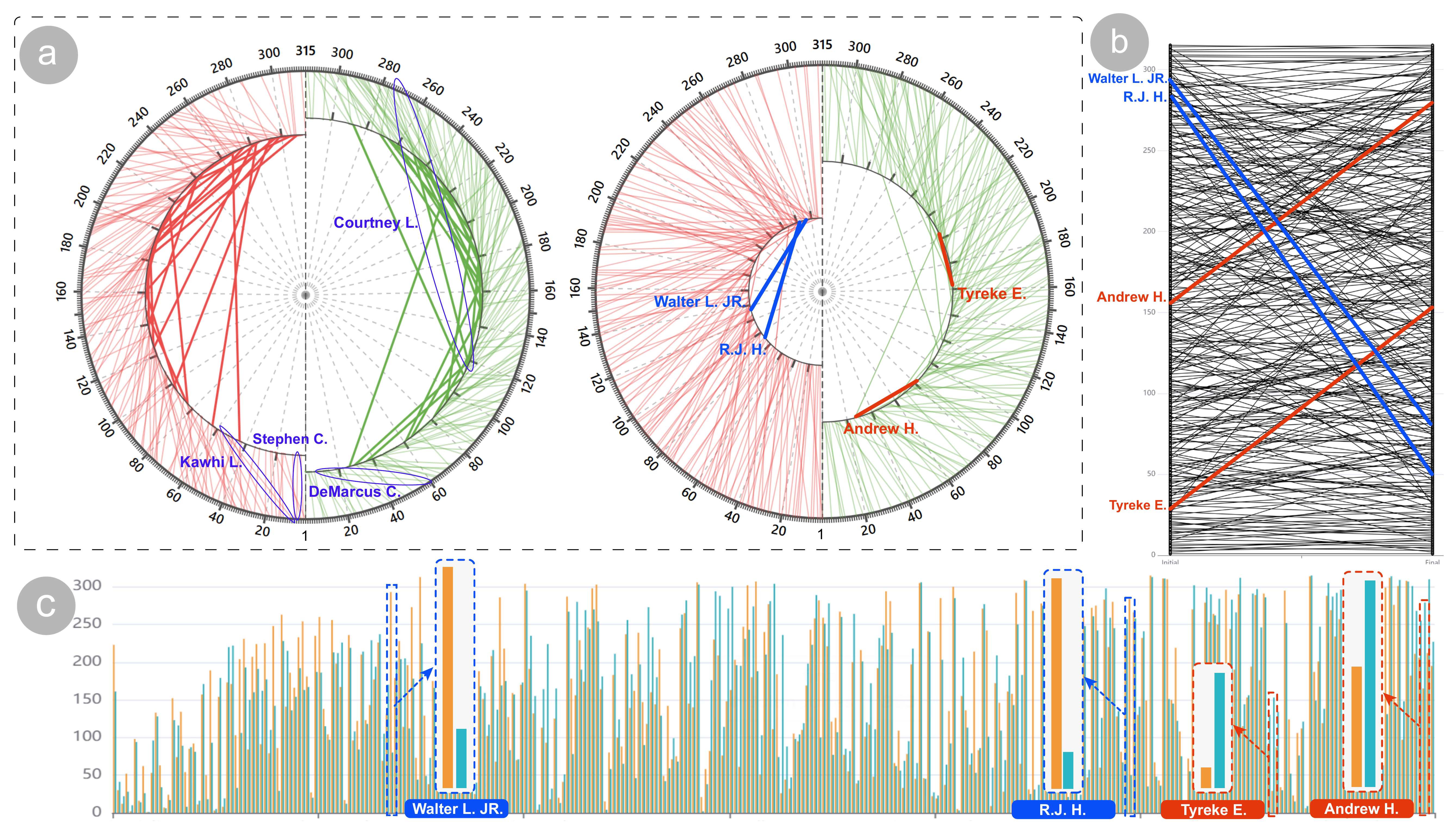}
  \caption{Comparison of a \toolName{}(a) and other existing visualization tools for state change comparison, i.e. slope graphs (b) and grouped bar charts (c), representing the Points per Games (PPG) changes of 321 players across two seasons from a basketball statistics. 
  The changes of PPG in two seasons before and after are indicated by intercepted line segment, line slopes and clustered bars` differences of (a), (b) and (c), respectively. 
(a)(left) accentuates players with top 30 PPG changes of the rising (left semi-circle, red) and dropping (right semi-circle, green) trends. 
  }
  \label{fig:teaser}
}

\abstract{
State change comparison of multiple data items is often necessary in multiple application domains, such as medical science, financial engineering, sociology, biological science, etc.
Slope graphs and grouped bar charts have been widely used to show a
``before-and-after'' story of different data states and indicate their changes.
However, they visualize state changes as either slope or difference of bars, which has been proved less effective for quantitative comparison. 
Also, both visual designs suffer from visual clutter issues with an increasing number of data items.
In this paper, we propose \toolName{}, a novel visual design to facilitate effective interactive comparison of state changes.
Specifically, a radial design is proposed to visualize the starting and ending states of each data item and the line segment length explicitly encodes the ``state change''.
By interactively adjusting the radius of the inner circular axis, \toolName{} can smoothly filter the large state changes and magnify the difference between similar state changes, mitigating the visual clutter issues and enhancing the effective comparison of state changes. 
We conducted a case study through comparing \toolName{} with slope graphs and grouped bar charts on real datasets to demonstrate the effectiveness of \toolName{}. 
}

\CCScatlist{
  \CCScatTwelve{Visual representation design}{Interaction}{State change comparison}{Radial visualization};
}

\begin{document}

\firstsection{Introduction}

\maketitle

% Comparison of state changes before and after 
State change comparison is one of the most commonly used methods for quantitative analysis~\cite{izquierdoevaluation}.
People need to compare the state changes of multiple data items and explore their initial and final states.
For example, NBA league will compare all players' progress and give the NBA Most Improved Player award to the player with the biggest progress compared to the regular season.
Also, daily new case numbers of coronavirus are used to evaluate the latest regional disease situation worldwide.
Visualization has been proved a powerful tool for data exploration. However, very few visualizations have been specifically designed for effectively visualizing and comparing multiple state changes.

% Visualization representation at the same time, is suitable to convey quantitative information of the comparison through graphical symbols.
% Therefore, the visual analytic of such state change comparison is of great importance and value in revealing the transition process and the comparison of item change quantities.

\begin{figure}[tbhp]
 \centering 
 \includegraphics[width=0.9\columnwidth]{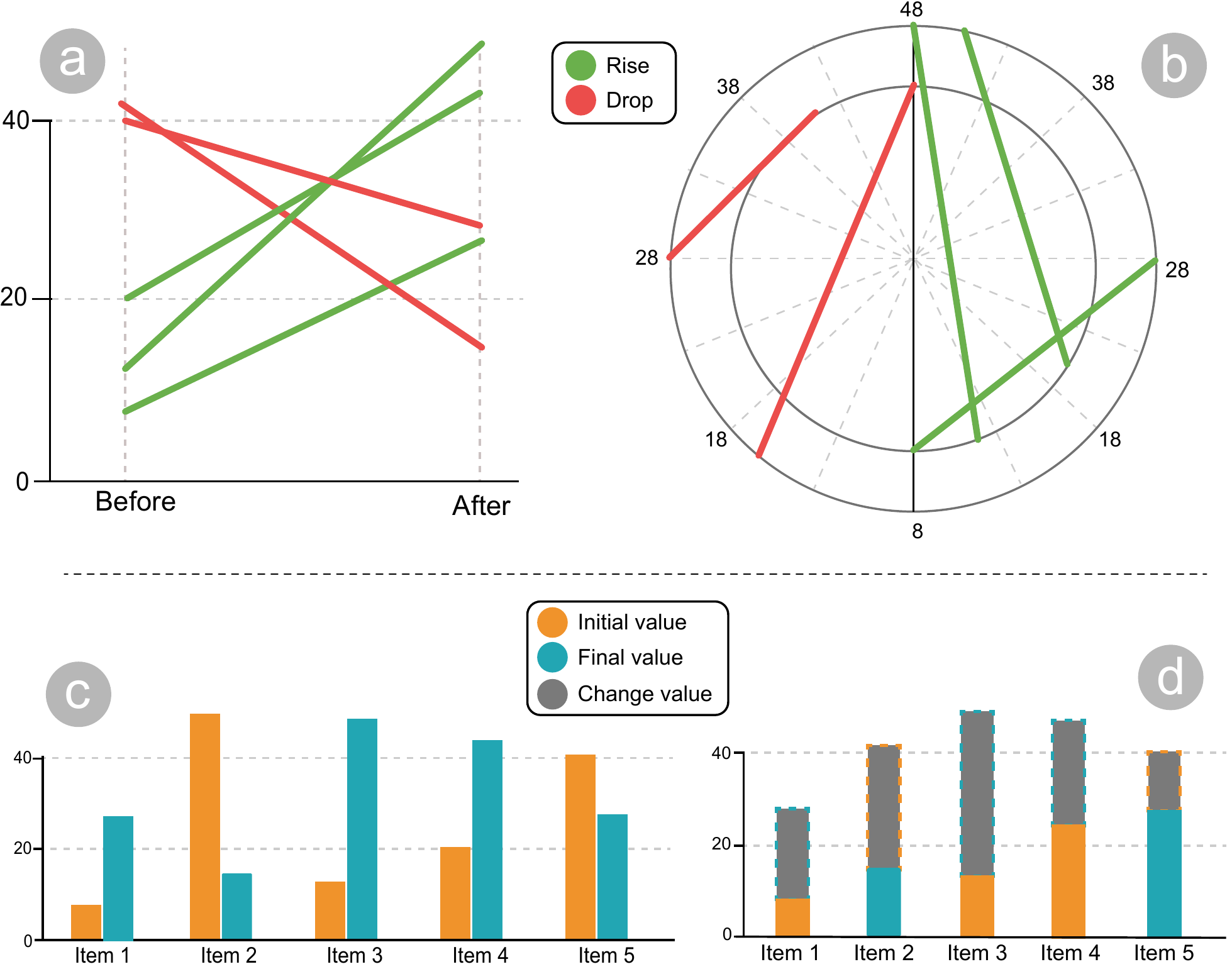}
 \caption{Existing visual designs and \toolName{} encoding the same data. \textbf{(a)} Slope graph \textbf{(b)} \toolName~ \textbf{(c)} Grouped bar chart \textbf{(d)} Stacked bar chart}
 \label{fig:2}
\end{figure}

According to our survey, slope graphs and grouped bar charts are often used to show and compare state changes 
due to their simplicity and accurate representation of the contexts (i.e. initial and final states). 
Slope graph \textbf{(Figure \ref{fig:2}a)} is a line graph that connects the initial and final states of each data item along two vertical axes, and the state changes are directly encoded by the line slopes.
Meanwhile, grouped bar charts \textbf{(Figure \ref{fig:2}c)} often use two adjacent bars to display the initial and final states of a data item, and multiple data items are shown along the x-axis. The state changes are implicitly encoded by the height differences within the grouped bars.

% However, there is very limited specifically-designed visualization type to show a two-state change story. According to our preliminary survey in Section \ref{sec:related_work}, slope graphs and grouped bar charts are dominant visual metaphors reflecting state change comparisons due to their simplicity and the nature of the highly accurate representation of the contexts (i.e. initial and final states). Slope graph \textbf{(Figure \ref{fig:2}(a))} is a line graph that connects dimension members across two points, and the quantity of changes is encoded by line slopes directly. As for grouped bar charts \textbf{(Figure \ref{fig:2}(c))}, bars are grouped by position for respective states before and after, and change quantities are encoded via illustrating the height differences within grouped bars.

Both slope graphs and grouped bar charts suffer from two major issues in supporting state change comparison.
Their first major limitation comes from the effectiveness of their visual encodings for state changes.
Slope graphs use \textit{slope} to indicate state changes. However, slope has been proven as a less accurate visual encoding channel than other encoding channels (e.g., \textit{length})~\cite{cleveland1987graphical,cleveland1986experiment,cleveland1985graphical}.
Grouped bar charts display state changes through the height differences of adjacent bars, but prior perception studies~\cite{burns2009modeling} have shown that people perform badly on the comparison of height differences of grouped bars.
The second major limitation of slope graphs and grouped bar charts is the visual clutter issue. 
With the increase of the visualized data items, slope graphs will suffer from severe visual clutters, as shown in Figure \ref{fig:teaser}b. 
For grouped bar charts, the bars will become thin and even difficult to recognize (Figure \ref{fig:teaser}c), and it becomes difficult to compare the height difference~\cite{doi:10.1198/106186002317375604}, especially such a comparison is distracted by various short and tall bars~\cite{talbot2014four}.
Thus, when the amount of data items exceeds the scalability of slope graphs and grouped bar charts, people are even forced to adopt data tables alternatively to represent the dataset~\cite{willers2017methods}.

% In graphical perception, studies have demonstrated that elementary code \textit{Slope} is a less accurate approach than \textit{Length} representation~\cite{cleveland1987graphical,cleveland1986experiment,cleveland1985graphical}, and people perform substantially badly on the comparison of height differences of grouped bars~\cite{burns2009modeling}. 
% Slope graphs, though, are capable of indicating more items over bar chart variants in limited screen space, the effectiveness of perception suffers from issues inherent to the mutual line overlapping (as shown in Figure \ref{fig:teaser}(b)). 
% Also, the bar separation for grouped bar charts would affect the comparison significantly when the bar amount exceeds a few dozens~\cite{doi:10.1198/106186002317375604} (as shown in Figure \ref{fig:teaser}(c)). 
% Prior work in the graphical perception community demonstrates that separated bar comparisons are more difficult than adjacent comparisons~\cite{elzer2006model,cleveland1984graphical}, especially when the separation contains the effects of short and tall distractor bars~\cite{talbot2014four}.
% Ultimately, according to our survey in Section \ref{sec:related_work}, researchers are forced to adopt data tables alternatively to represent data set when data amounts exceed the scalability of preceding visual metaphors for change representation, which exerts improved and scalable requirements on visualization representation methods.

To address the above two major issues,
we propose \toolName{} (\textbf{Figure \ref{fig:2}b}), a novel radial visual design to facilitate effective comparison of state changes across multiple data items.
Specifically, it allows a context-aware change comparison with the length of the line segments, which is proved as a more effective channel over slopes and bar differences. 
Also, with an increasing number of data items, \toolName{} facilitates the change representation to support quick and smooth accentuation of large state changes and effective comparison among relatively similar state changes via introducing intuitive interactions.

% which are generally drawn from the inner circular axis to the outer circular axis. The layout of left and right semi-circles is used to visualize state changes with negative and positive respectively. 

% By introducing intuitive interactions, \toolName{} facilitates the change representation to support better comparison via improved graphical perception methods.

% Specifically, as the inner axis shrinks inward, data items with relatively larger change quantities will be smoothly filtered out and accentuated inside the inner axis. Also, \toolName{} supports the comparison of similar state changes interactively. Both techniques perform well with an increasing number of data items.

% Through user interactions, \toolName{}  supports the accentuation of items with large change quantities if viewers intend to dismiss items with normal state changes. Also, a useful feature is proposed to facilitate the change comparison interactively, which addresses the challenge of inaccurate comparison for existing visualization tools (i.e. slope graphs and grouped bar charts).

We have released our approach as a publicly-available JavaScript library called \textit{interceptgraph}
% \url{https://www.npmjs.com/package/interceptgraph}.
~\footnote{\url{https://www.npmjs.com/package/interceptgraph}}, which enables a quick comparison of multiple state changes.

The major contributions can be summarized as follows:

\begin{itemize}
    \item We propose \toolName{}, a novel radial visualization tool for effective visual comparison of multiple state changes, which leverages the line segment length to directly encode the state changes, and intrinsically enables smooth interactions for filtering large state changes of user interest and amplifying the difference of similar state changes for an accurate comparison.

    \item We conducted case studies and compared \toolName{} with slope graphs and grouped bar charts to evaluate its performance. The results demonstrate the usefulness and effectiveness of \toolName{}.  
    
    % \item We present simple interactive methods for 
    % comparison with improved graphical perception, and quick accentuation feature for items with large state changes.
    
    % \item We investigate the effectiveness of the proposed tool \toolName through a case study in Section \ref{sec:case_study}, in comparison with the most commonly used visualization tools.
\end{itemize}

 % include \maketitle
\section{Preliminary Survey}

% To the best of our knowledge, 
There are few prior studies specifically investigating the visualization of state changes.
Thus, to identify what visualizations have been applied to visualizing state changes,
% verify our potential targets,
% (i.e., slope graphs and grouped bar charts), 
we conducted a preliminary survey to determine the mostly used visualization types for statistical change comparison.
Following the methodology used by Segel and Heer~\cite{segel2010narrative}, we first gathered figures from existing research papers that need to compare multiple state changes. 
We used the permutation of ``state'', ``change'', ``comparison'' as search keywords and manually harvested 100 top query results from Google Scholar. 
Since each study may include multiple figures for state change comparison, we further split them into 156 individual figure units.
We then categorized all figure units into the five main groups (Table \ref{table:2.1.1}) introduced by Borkin et al.~\cite{borkin2013makes}. 
Note that the \textit{Heatmap} category is designed to visualize changes with regard to spatial information such as the physical position, which is beyond the scope of our study and thus excluded from our survey.

\begin{table}[hbtp]
\begin{center}
\begin{tabular}{@{}c|c|r@{}}
\toprule
\multicolumn{2}{c|}{Category}               & \multicolumn{1}{c}{Percentage} \\ \midrule
\multirow{2}{*}{Bar}    & Grouped Bar Chart & 44.9\%                         \\
                        & Stacked Bar Chart & 0.6\%                          \\ \midrule
Line                    & Slope Graph       & 28.2\%                         \\ \midrule
\multirow{2}{*}{Circle} & Pie Chart         & 3.8\%                          \\
                        & Donut Chart       & 0.6\%                          \\ \midrule
Grid \& Matrix          & Heatmap           & 19.9\%                         \\ \midrule
Points                  & Scatter Plot      & 1.9\%                          \\ \bottomrule
\end{tabular}
\end{center}
\caption{Categories of figure units and respective percentages collected from related research papers.
}
\label{table:2.1.1}
\end{table}

\section{related work}
\label{sec:related_work}

The related work of this paper can be categorized into two groups: state change visualization 
and 
radial visual design.

% \subsection{Visualization Tools for State Change Representation}
\subsection{State Change Visualization}

According to the preliminary survey shown in Table \ref{table:2.1.1}, we finally decide to target grouped bar charts and slope graphs due to their dominance in our harvested data set. 

Slope graph~\cite{tufte1985visual} (Figure \ref{fig:2}a) is an appropriate visual design when the nature of the task is to compare state changes across items based on comparing their line slope in time. A positive value of the slope implies that the dependent variable increases, while a negative value implies that the variable decreases. 
Grouped bar chart~\cite{beniger1978quantitative} (Figure \ref{fig:2}c) is another approach to display state changes with the context of initial and final values, which encodes the initial and final values by respective categorical bars within each group. 
However, distractors between two target bar groups inevitably affect graphical perception when the amount of items exceeds its scalability~\cite{talbot2014four, doi:10.1198/106186002317375604}, 
grouped bar chart is the most common method to show state changes.

Stacked bar chart~\cite{donnelly2009humongous} (Figure \ref{fig:2}d) is the most straightforward solution when we previously interviewed domain experts, which 
indicates change counts by the stacked sub-bars on lower sub-bars denoting a context state.
% extends classic bar charts via stacking sub-bars for changes onto lower sub-bars denoting a context state, 
% while the overall bar height indicates the other state
However, if the data set includes data items of both rise and drop trends, the representation may suffer from visual complexity with an increasing number of items, which would significantly affect the human perception. Also, viewers can not determine relative bar height accurately on such unaligned bar chart variants~\cite{cleveland1987graphical}. As shown in Table \ref{table:2.1.1}, researchers rarely utilize this visualization type to compare state changes.

% Variants for bar charts
%  There are still other ways to modify bar charts. 
%  For instance, cut-off bars, linear bar charts with scale break~\cite{1992Graphic}, and scale-stack bar charts~\cite{2013Scale} have been proposed to solve the scalability problems. However, they are all using a vertical data representation that covers a large value range, and generally cannot improve the scalability of the horizontal bar arrangement.

In this paper, the state changes are encoded by the \textit{lengths} of different line segments,
% of different \textit{lengths}, 
which is more accurate than \textit{height difference} (for grouped bar charts) and \textit{slopes} (for slope graphs)~\cite{cleveland1987graphical,cleveland1986experiment,cleveland1985graphical}. 
% which is an more accurate elementary code over \textit{unaligned lengths} (for grouped bar charts) and \textit{slopes} (for slope graphs)~\cite{cleveland1987graphical,cleveland1986experiment,cleveland1985graphical}. 
Also, intuitive interactions are enabled in \toolName{} to support a comparison of state changes with better graphical perception.

\subsection{Radial Visual Design}

% \yong{pls restructure the related work by following such logic flow:
% 1. One sentence summarizing the prior studies and their taxonomy.
% 2. Introduce each type of work and representative papers.
% 3. Another paragraph to summarize the major difference or highlights of our work.
% }

Visual representations of data that are based on circular shapes are referred to as radial visualizations~\cite{burch2014benefits}.
Draper et al.~\cite{draper2009survey} provided a comprehensive survey on radial visualization and categorize it into three visual themes: \textit{Polar Plot}, \textit{Space Filling} and \textit{Ring Based}. 
The earliest use of a radial display in statistical graphics was the pie chart, which was proposed in William Playfair's 1801 treatise, the Statistical Breviary~\cite{playfair1801statistical}. 
After that, radial visualization is becoming an increasingly pervasive metaphor in information visualization. 
Radviz~\cite{grinstein2002information} is a typical radial visualization-based approach to cluster multidimensional data. Hacıaliefendioğlu et al.~\cite{hacialiefendiouglu2020co} developed a radial technique that allows elaborate visualization of the interplay between different violence types and subgroups. Additionally, prior studies further discussed the strengths and weaknesses of radial visualization through various methodologies~\cite{diehl2010uncovering,goldberg2011eye}.

According to the taxonomy presented by Draper et al.~\cite{draper2009survey}, \toolName{} belongs to the subtype \textit{Connected Ring Pattern} under \textit{Ring Based}. Accordingly, \toolName{} preserves the advantages of radial visualization and further extends static radial methods via flexible interactions, making it available to compare items more accurately and effectively.

% \section{visualization method}

\section{Visual Design}

We describe the composition of \toolName{}, the approach of adjusting the radius of the inner circular axis, and the user interaction.

\subsection{Visualising an Intercept Graph}
\label{subsec:basic_method} 

\textbf{Intercept Graph} uses line segments to facilitate the comparison of state changes across multiple data items. 
% is so named because intercepted chord length is used for state change representation and comparison. We propose a method of visualizing the change quantity with a chord length mapping. 
The inner and outer circular axes are used to locate ``initial'' and ``final'' states respectively. Note that \toolName{} is not an intact dual-circle design, since the left and right semi-circular axis are separated apart intrinsically, which are used to visualize data items with drop and rise trends of state values respectively.

\begin{figure}[t]
 \centering 
  \begin{mdframed}
 \includegraphics[width=\columnwidth]{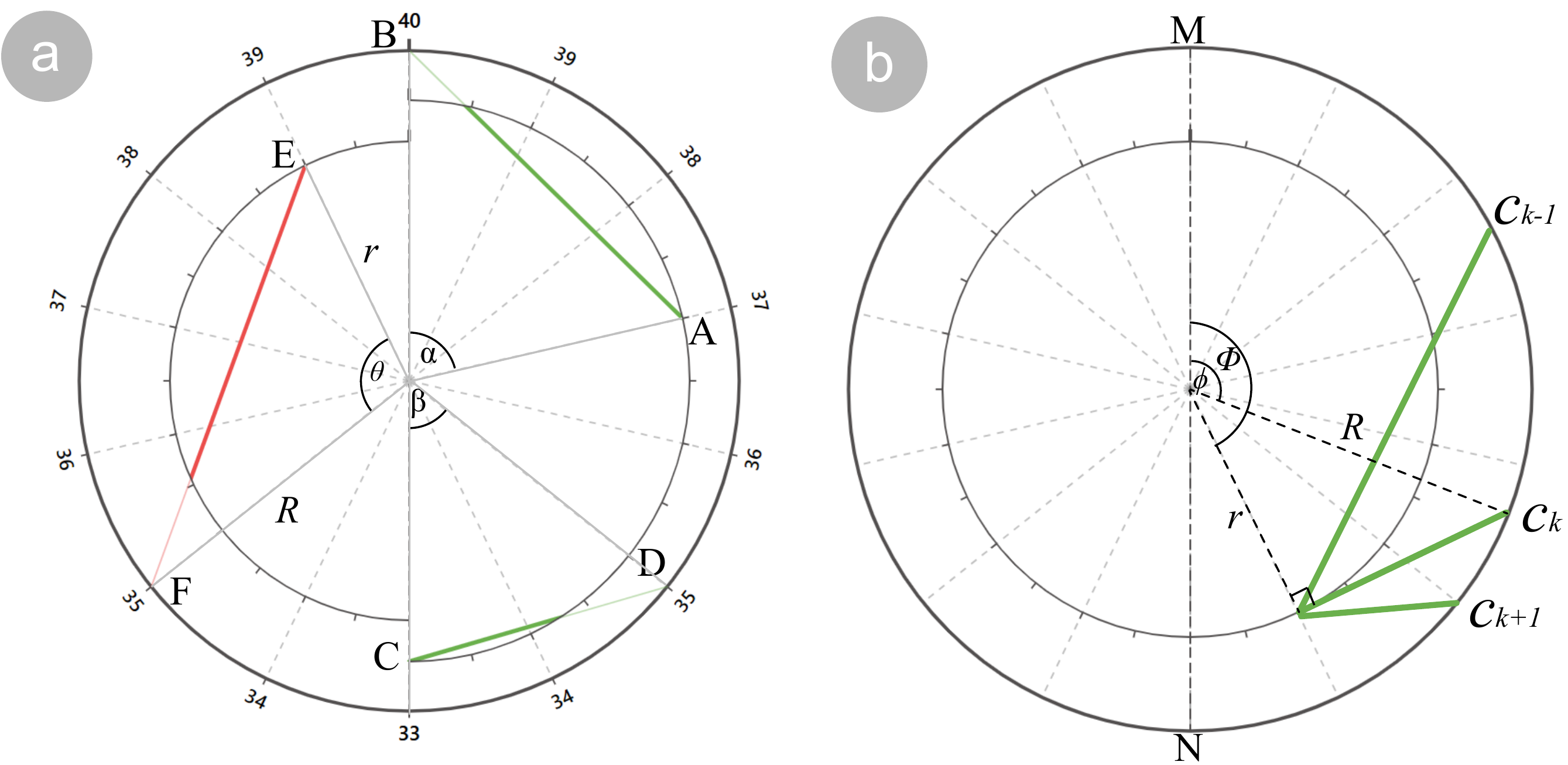}
 \end{mdframed}
 \caption{(a) An example showing that state changes are linearly encoded by central angles. (b) Analytic geometry diagram of \toolName{} for the calculation of the radius of inner circular axis.
 }
 \label{fig:4}
\end{figure}

\modifY{\textbf{Line segments} (e.g., \textit{Line AB, Line CD, Line EF} in Figure \ref{fig:4}a) are a set of lines generally drawn from the inner circular axis to the outer circular axis, which are used to implicitly encode the change quantity of each item. The central angle between radii representing initial and final values is proportional to the state changes as the scale of both inner and outer circular axes are linearly distributed. For example, suppose that there are two data items. One data item changes from 33 to 35 and the other from 37 to 40. Then the ratio of the central angles of Intercept Graph is 3:2 as shown in the angles $\alpha$ and $\beta$ subtended to line segments \textit{AB} and \textit{CD} in Figure \ref{fig:4}a. Also, following the Lows of Cosines, the line segment $c$ is determined as follows in terms of $\theta$:}

\begin{equation}
% r(k) = R \cdot \cos (\vert \Theta_{k+1} - \theta_{k+1} \vert)
c = \sqrt{r^2+R^2-2 \cdot r \cdot R \cdot \cos{\theta}}
\label{equ:1}
\end{equation}

\modifY{where constants $r, R$ denote the radii of the inner and outer circular axis respectively (as shown in the line segment \textit{EF} in Figure \ref{fig:4}a). $\theta \in [0,\pi]$ denotes the central angle subtended to the line segment. Equation \ref{equ:1} is monotonic increasing in terms of $\theta$, which indicates that the central angle of \toolName{} is correlated positively with the line segment length. So, according to the two conclusions illustrated above, the line segment length is positively correlated with the change quantity.}

% \textbf{Chord length} is used to implicitly encode the change quantity. \modifY{The central angle between radii representing initial and final values is proportional to the state changes as the scale of both inner and outer radial axes are linearly distributed. For example, suppose that there are two data items. One data item changes from 33 to 35 and the other from 37 to 40. Then the ratio of the central angles of Intercept Graph is 3:2 as shown in Figure \ref{fig:4}a. Also, following the first equation illustrated in Problem 2 of~\cite{bruce1943approximations}, the central angle is correlated positively with the chord length. In conclusion, the chord length is positively correlated with the change quantity.}
% \yong{Pls check my side comments.}

\begin{figure}[t]
 \centering 
 \begin{mdframed}
 \includegraphics[width=\columnwidth]{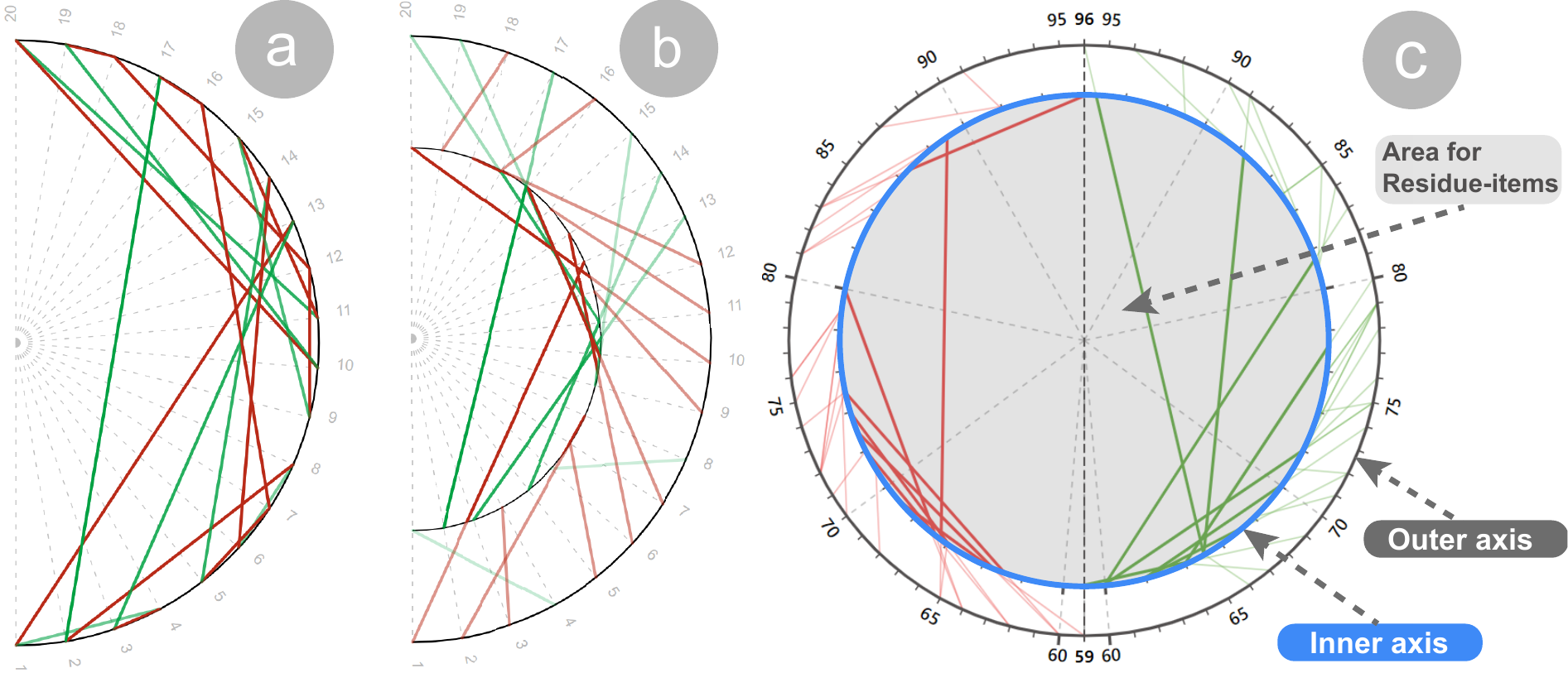}
 \end{mdframed}
 \caption{Alternative designs of \toolName. (a) A draft with lines in the same semi-circular axis. (b) Extending (a) by introducing the inner circular axis for item filtering. (c) 
 The final visual design.
%  An intact circle with separated parts.
 }
 \label{fig:3}
\end{figure}

\textbf{Axis range} is determined by the minimum and maximum of the ``initial'' and ``final'' states of all the data items.
% which can improve the graphical perception and aesthetics. 
\modify{
With such a setting, \toolName{} can have more space to highlight the state changes, facilitating an easy comparison of different state changes. 
% Specifically, the range setting yields benefits of the user interaction for change comparison because the chord lengths of items would be longer, making them more intuitively to be compared.
As shown in Figure \ref{fig:3}c, both the left and right parts of \toolName{} have a fixed radius of outer circular axis and adjustable radius for the inner circular axis.
% a fixed outer axis and an adjustable inner axis for a better quantity comparison performance via chord segments intercepted by the inner axis.
} 
% Thus, an introduction of items with an intercepted part within the inner axis, we call these items \textit{residue-items}, is further organized.
% \yong{pls check my side comments.}

\textbf{Residue-items} are the remaining data items indicated by the line segments who intersect with the inner circular axis, as shown by the line segments with a bold portion in Figure \ref{fig:3}c.
% are used to indicate the filtered items, whose responding chords have bold segments intercepted by the inner axis (as shown in items with bold segments in Figure \ref{fig:3}c). 
The set of \textit{residue-items} varies according to the adjustment of the radius of the inner circular axis, which serves as a filter which keeps only the items with a relatively large change. \modify{More specifically, the smaller the inner circular axis, the fewer \textit{residue-items}. Otherwise, more data items with relatively small state changes will also be kept.}

\textbf{Alternative designs:}
Before we come up with the current design, we also considered several alternative designs (Figures \ref{fig:3}a and b).
% of \toolName{} are shown in Figure \ref{fig:3}. 
% A draft of \toolName, the half circle design (Figure \ref{fig:3}(a)), has multiple items with initial and final state locating on the same semi-circular axis.
Figure \ref{fig:3}a can visualize the initial and final states of multiple data items, but they cannot support interactively filter data items with a large change.
% includes the primary required elements. 
Figure \ref{fig:3}b enables interactive filtering of \textit{residue-items}, but still suffers from serious visual clutter.
% with interactions. 
% Figure \ref{fig:3}c is preferred, 
\toolName{} is preferred, 
as it mitigates the visual clutter by plotting increasing and decreasing data items in the left and right circular axes, respectively.
% \yong{pls check if I made it wrong.}
% by delimiting items with different change signs by separated circular axes.
% The visual designs shown in Figure \ref{fig:3}a visualizes the initial and final states of multiple data items. However, they are not preferred as no residue-item is supported without interactions. We introduce the inner axis (Figure \ref{fig:3}b) to fix the problem, but visual clutter still exists. Thus, we delimited items with different change signs by separated circular axes (Figure \ref{fig:3}c) to improve the visual clutter.

% We implemented the proposed visual design with web-based SVG elements. A library called \textit{interceptgraph}~\footnote{https://www.npmjs.com/package/interceptgraph} supports instant build of Intercept Graph for developers or ordinary people.

% \yong{Reach here.}

\subsection{Radius of the Inner Circular Axis}
% \textit{vs.} Residue Items
\label{subsection:4.2}

\modify{
With the decrease of radius of the inner circular axis, all the data items with a smaller state change will be excluded from the residue items, i.e., \textit{the state changes of all the \textit{residue-items} are always larger than those excluded from the \textit{residue-items}}.
Figure \ref{fig:4}b provides an intuitive illustration for this.
As introduced in Section \ref{subsec:basic_method}, the line segment length is positively correlated with the change quantity.
Suppose we decrease the inner circular axis outward until it is tangent to \textit{Line} $c_{k}$, which corresponds to the data item with $k$-th largest state changes.
Then, \textit{Line} $c_{k-1}$ (representing the $k-1$ largest state changes) should always be included in the \textit{residue-items}, while \textit{Line} $c_{k+1}$ (representing the $k+1$ largest state changes) is already excluded from the \textit{residue-items}.
}

\modifY{
Given the above properties of the radius of the inner circular axis, users can interactively adjust the radius of the inner circular axis to focus on the data items with higher state changes.
Also, we provide an automated way to help users quickly filter the data items with top-$k$ state changes by automatically determining the corresponding radius of inner axis.
% According to Equation \ref{equ:1}, the radius can be calculated as follows:
As shown in Figure \ref{fig:4}b,
the corresponding radius of inner circular axis $r$ can be calculated as follows:
% \begin{equation}
% r(k) = R \cdot \cos (\vert \Phi_{k} - \phi_{k} \vert)
% \label{equ:2}
% \end{equation}
\begin{equation}
r = R \cdot \cos (\vert \Phi - \phi \vert)
\label{equ:2}
\end{equation}
where $\Phi,\phi \in [0, 2\pi]$ denote the angles between the vertical separating line \textit{MN} and the corresponding radii indicating the initial and final states of the data item with the $k$-th largest state change.
% where $\Phi \in [0, 2\pi]$ denotes the angle between the vertical separating line \textit{MN} and $r$, and $\phi \in [0, 2\pi]$ denotes the angle between the vertical separating line \textit{MN} and $R$.
}

\subsection{User Interaction}

% Adjusting the inner axis radius can result in two aspect results: 
% selecting residue-items based on their absolute value of change quantities, and controlling the ratio of the length of residue-items` intercepted chord segments.

% As the inner axis expanding outward, items with relatively small change quantity will change into residue-items. The whole set of items will become residue-items eventually when the inner axis radius reaches the upper limit (outer circular axis radius). Otherwise, all items will be filtered when the inner radius is equivalent to zero.

The user interaction extends \toolName{} from static radial visualization. Specifically, two features called \textit{large change accentuation} and \textit{close change magnification} are proposed to supports more advanced features over the basic nature plotting change counts.

\textbf{Large change accentuation} allows quick filtering for the data items with large state changes of user interests. For example, through shrinking the radius of the inner circular axis, items with larger change counts would be more likely to be filtered (the flow is shown from Figure \ref{fig:3}a to Figure \ref{fig:3}b). Otherwise, all data items will turn into \textit{residue-items} when the inner circular axis radius is equivalent to that of the outer circular axis. This feature performs well with an increasing number of data items.

\textbf{Close change magnification} enhances the human graphical perception of state change comparison through amplifying the difference of similar change quantities interactively (as shown in pairwise items highlighted in dark blue and crimson in Figure \ref{fig:teaser}). Through shrinking the inner circular axis inward, the ratio of pairwise line segments will be magnified, which makes the comparison of relative state changes more effective.

\section{case studies}
\label{sec:case_study}

We conduct a case study on a basketball dataset to demonstrate the effectiveness of our proposed visual design. It contains 321 NBA player statistics, who are active players in both Season 2018 and 2019. We adopt the application as evaluating the progress of players is of great importance in the league, which is attributed to the foundation of the annual award Most Improved Player~\cite{martinez2019method}. Following the methodology proposed by Dumitrescu et al.~\cite{dumitrescu2000evolutionary}, we use a rank-based statistical category Points per Game (abbreviated as PPG) instead of row data to address the discrepancy between players' performance and the highly-aggregated PPG records.

The preceding conventional designs, such as slope graph (Figure \ref{fig:teaser}b), shows a PPG trend story by connecting two PPG states of Season 2018 and 2019, while another target design grouped bar charts (Figure \ref{fig:teaser}c) plots items with 321 clustered bars. 
Apparently, both designs have limitations to visualize PPG changes. First, they encode changes by ineffective visual channels. For slope graphs (Figure \ref{fig:teaser}b), line slopes across different players are difficult to be compared, especially there are distractors between two target lines. Also, as shown in the detailed view of Figure \ref{fig:teaser}c, bar height differences indicating PPG changes can not be perceived effectively. 
Furthermore, both designs are beyond their respective visual scalability to plot over 300 items. Specifically, for grouped bar charts, the perception suffers from the visual clutter in terms of the narrow width of bars and a variety of distractor bars. For slope graphs, serious line overlapping makes it hard to distinguish different lines and compare line slopes.
% Apart from the perception issues of state changes, as shown in Figure \ref{fig:teaser}(b) and Figure \ref{fig:teaser}(c), the comparisons of state change quantities of pairwise items highlighted by dark blue and orange are difficult because of other dense distractors and inaccurate change encoding.

On the contrary, the proposed visual design \toolName{} improves state comparisons in terms of better graphical perceptions. Figure \ref{fig:teaser}a-left is used to visualize the top 30 players of rise and drop PPG changes by setting the \textit{residue-item} number to 30 interactively. It is clear that the 30 \textit{residue-items} for both rise and drop trends are arranged sparsely within the inner circular axes, which mitigates visual clutter issues significantly. If, instead, the user is interested in sets of top 10 candidates of MIP selection, simply setting the \textit{residue-item} number to 10 will fulfill the needs (Figure \ref{fig:teaser}a-right).
Also, based on the length mapping of state changes, it is apparent to recognize that Kawhi Leonard (Line \textit{Kawhi L.}) has a larger PPG progress than Stephen Curry (Line \textit{Stephen C.}), \modify{both of which are plotted as red line segments due to the decrease of the rank values (e.g., from the third to the first), which actually indicates an improvement of their PPGs.}
% \yong{pls check if I change your idea.}
Also, the PPG of Courtney Lee (Line \textit{Courtney L.}) drops much more than that of DeMarcus Cousins (Line \textit{DeMarcus C.}) due to the longer line segment length.
% Apparently, \toolName{} enhances the ability of viewers to quickly filter players with large PPG changes interactively, making it more intuitive to pick MIP candidates via a more time-saving and accurate method.

More interesting findings can be revealed by \toolName. Here, we introduce a statistical measure \textit{percentage difference}, according to a prior study~\cite{cole2017statistics}, to reflect differences of two lengths of intercepted line segments. 
% As previous stated, our approach improves the graphical perception of player's PPG difference comparison through magnifying the ratio of intercepted chord length interactively. 
As shown in Figure \ref{fig:teaser}b and Figure \ref{fig:teaser}c, Line \textit{Walter L. JR.} and Line \textit{R.J. H.} (highlighted in dark blue annotations) have a percentage difference of slopes and bar height differences of 8.9\% (213 and 234 places risen respectively) due to the linear visual mapping.
However, our approach (Figure \ref{fig:teaser}a-right) magnifies the length ratio of intercepted line segments to 18.3\% (100.9 pixels to 123.4 pixels) through image software measurement. 
Another pair of target players \textit{Andrew H.} and \textit{Tyreke E.} for drop trend of PPG ranking (highlighted in crimson annotations) have the percentage differences of 8.1\% (114 and 124 places dropped) and 19.2\% (54.8 pixels to 67.8 pixels) for two preceding designs and our \toolName{} respectively. It is clear that both results magnify the original linear mapping over two times, which makes the original linear mapping apparent enough to make judgments.
% The comparison in \toolName{} is apparent as it amplifies the length ratio interactively and leverages a more accurate elementary code to indicate state changes.

\section{conclusion}
\label{sec:conclusion}

% repeat advantages
In this paper, we present a novel visual design \toolName{} for context-aware comparison of state changes. \modify{Instead of focusing on visualizing the exact change quantities, \toolName{} is mainly designed for facilitating the \textit{comparison} of state changes across multiple data items via more effective interaction.
We compared \toolName{} with widely-used established tools (i.e., slope graphs and grouped bar charts)}. A case study on a two-season basketball dataset shows that our design can quickly filter large state changes and amplify the difference of similar state changes for an accurate comparison through smooth interactions.
In future work, we plan to conduct more case studies and user studies on real datasets to further evaluate the effectiveness of \toolName{}. Also, it would be interesting to explore how to \textit{automatically} determine the optimal default inner circular radius for more efficient comparison of state changes.
% \yong{Shaolun, pls check if you are fine with it.}

% However, several aspects of the proposed \toolName{} approach still need future work. First, the radius of the inner radial axis is manually tuned for the best observation effect. Other methods for radius determination have not yet be investigated, e.g., methods of determining the level of state changes through the inner axis. Also, our case study has demonstrated that \toolName{} is capable to better visualize items with large change quantities. However, it remains possible if the goal is the analysis of items with small changes instead through transferring the direction of the inner and outer radial axis into the opposite. The point is to demonstrate the usefulness and geometric nature of this variant, which is left as future research.

\acknowledgments{
This research was supported by the Singapore Ministry of Education (MOE) Academic Research Fund (AcRF) Tier 1 grant (Grant number: 20-C220-SMU-011).
S. Ruan was partially supported by the National Natural Science Foundation of China (Grant No. 61872066 and U19A2078), and the Science and Technology project of Sichuan (No. 2020YFG0056).}

\bibliographystyle{abbrv-doi}

\bibliography{template}

\begin{thebibliography}{10}

\bibitem{beniger1978quantitative}
J.~R. Beniger and D.~L. Robyn.
\newblock Quantitative graphics in statistics: A brief history.
\newblock {\em The American Statistician}, 32(1):1--11, 1978.

\bibitem{borkin2013makes}
M.~A. Borkin, A.~A. Vo, Z.~Bylinskii, P.~Isola, S.~Sunkavalli, A.~Oliva, and
  H.~Pfister.
\newblock What makes a visualization memorable?
\newblock {\em IEEE Transactions on Visualization and Computer Graphics},
  12(19):2306--2315, 2013.

\bibitem{burch2014benefits}
M.~Burch and D.~Weiskopf.
\newblock On the benefits and drawbacks of radial diagrams.
\newblock In {\em Handbook of human centric visualization}, pp. 429--451.
  Springer, 2014.

\bibitem{burns2009modeling}
R.~Burns, S.~Carberry, and S.~Elzer.
\newblock Modeling relative task effort for grouped bar charts.
\newblock In {\em Proceedings of the Annual Meeting of the Cognitive Science
  Society}, vol.~31, 2009.

\bibitem{cleveland1985graphical}
W.~S. Cleveland and R.~McGill.
\newblock Graphical perception and graphical methods for analyzing scientific
  data.
\newblock {\em Science}, 229(4716):828--833, 1985.

\bibitem{cleveland1986experiment}
W.~S. Cleveland and R.~McGill.
\newblock An experiment in graphical perception.
\newblock {\em International Journal of Man-Machine Studies}, 25(5):491--500,
  1986.

\bibitem{cleveland1987graphical}
W.~S. Cleveland and R.~McGill.
\newblock Graphical perception: The visual decoding of quantitative information
  on graphical displays of data.
\newblock {\em Journal of the Royal Statistical Society: Series A (General)},
  150(3):192--210, 1987.

\bibitem{cole2017statistics}
T.~J. Cole and D.~G. Altman.
\newblock Statistics notes: What is a percentage difference?
\newblock {\em BMJ: British Medical Journal (Online)}, 358, 2017.

\bibitem{diehl2010uncovering}
S.~Diehl, F.~Beck, and M.~Burch.
\newblock Uncovering strengths and weaknesses of radial visualizations---an
  empirical approach.
\newblock {\em IEEE Transactions on Visualization and Computer Graphics},
  16(6):935--942, 2010.

\bibitem{donnelly2009humongous}
R.~Donnelly and W.~M. Kelley.
\newblock {\em The Humongous Book of Statistics Problems: Nearly 900 Statistics
  Problems with Comprehensive Solutions for All the Major Topics of
  Statistics}.
\newblock Penguin, 2009.

\bibitem{draper2009survey}
G.~M. Draper, Y.~Livnat, and R.~F. Riesenfeld.
\newblock A survey of radial methods for information visualization.
\newblock {\em IEEE Transactions on Visualization and Computer Graphics},
  15(5):759--776, 2009.

\bibitem{dumitrescu2000evolutionary}
D.~Dumitrescu, B.~Lazzerini, L.~C. Jain, and A.~Dumitrescu.
\newblock {\em Evolutionary computation}.
\newblock CRC press, 2000.

\bibitem{doi:10.1198/106186002317375604}
S.~G. Eick and A.~F. Karr.
\newblock Visual scalability.
\newblock {\em Journal of Computational and Graphical Statistics},
  11(1):22--43, 2002. doi: {{%
10\hspace{.1pt}\discretionary{.}{%
}{.}\hspace{.4pt}1198\discretionary{/}{%
}{/}106186002317375604}}


\bibitem{grinstein2002information}
U.~Fayyad, G.~G. Grinstein, and A.~Wierse.
\newblock Information visualization in data mining and knowledge discovery,
  2001.

\bibitem{goldberg2011eye}
J.~Goldberg and J.~Helfman.
\newblock Eye tracking for visualization evaluation: Reading values on linear
  versus radial graphs.
\newblock {\em Information Visualization}, 10(3):182--195, 2011.

\bibitem{hacialiefendiouglu2020co}
A.~Hac{\i}aliefendio{\u{g}}lu, S.~Y{\i}lmaz, M.~Koyut{\"u}rk, and G.~Karakurt.
\newblock Co-occurrence patterns of intimate partner violence.
\newblock In {\em Proceedings of the Pacific Symposium}, pp. 79--90. World
  Scientific, 2020.

\bibitem{izquierdoevaluation}
E.~M. Izquierdo, C.~G. Mart{\'\i}n, A.~A. Hidalgo, and M.~L. Saavedra.
\newblock Evaluation of a change detection method based on joint
  spatial-spectral information.

\bibitem{martinez2019method}
J.~A. Martinez.
\newblock A method to evaluate the most improved player in basketball based on
  a non-linear difficulty curve.
\newblock {\em Khel Journal}, 2019.

\bibitem{playfair1801statistical}
W.~Playfair.
\newblock {\em The statistical breviary}.
\newblock Wallis, 1801.

\bibitem{segel2010narrative}
E.~Segel and J.~Heer.
\newblock Narrative visualization: Telling stories with data.
\newblock {\em IEEE Transactions on Visualization and Computer Graphics},
  16(6):1139--1148, 2010.

\bibitem{talbot2014four}
J.~Talbot, V.~Setlur, and A.~Anand.
\newblock Four experiments on the perception of bar charts.
\newblock {\em IEEE Transactions on Visualization and Computer Graphics},
  12(20):2152--2160, 2014.

\bibitem{tufte1985visual}
E.~R. Tufte.
\newblock The visual display of quantitative information.
\newblock {\em The Journal for Healthcare Quality (JHQ)}, 7(3):15, 1985.

\bibitem{willers2017methods}
J.~Willers.
\newblock {\em Methods for extracting data from the Internet}.
\newblock PhD thesis, Iowa State University, 2017.

\end{thebibliography}
\end{document}